\documentclass[aps,prd,preprintnumbers,preprint,superscriptaddress,showpacs,floatfix]{revtex4-1}

\usepackage{epsfig}
\usepackage{bm}
\usepackage{amssymb}
\usepackage{amsmath}
\usepackage{color}
\usepackage{subfigure}
\newcommand{\be}{\begin{equation}}
\newcommand{\ee}{\end{equation}}
\newcommand{\bea}{\begin{eqnarray}}
\newcommand{\eea}{\end{eqnarray}}

\begin{document}

\title{Azimuthal asymmetries in high-energy collisions of protons with
  holographic shockwaves}
\author{Jorge Noronha}
\email{noronha@if.usp.br}
\affiliation{Instituto de F\'{\i}sica, Universidade de S\~{a}o Paulo,
  C.P. 66318, 05315-970 S\~{a}o Paulo, SP, Brazil}
\author{Adrian Dumitru}
\email{adrian.dumitru@baruch.cuny.edu}
\affiliation{Department of Natural Sciences, Baruch College, CUNY,
17 Lexington Avenue, New York, NY 10010, USA}

\begin{abstract}
Large azimuthal quadrupole and octupole asymmetries have recently been
found in p+Pb collisions at the LHC. We argue that these might arise
from a projectile dipole scattering off fluctuations in the target
with a size on the order of the dipole. In a holographic scenario, parity even angular moments
$v_{2n}$ are generated by the real part of the light-like Wilson loop
due to the contribution from the background metric to the Nambu-Goto
action. On the other hand, parity odd moments $v_{2n+1}$ must arise from the imaginary part of
the light-like Wilson loop, which is naturally induced by a fluctuating Neveu-Schwarz 2-form.
\end{abstract}

\maketitle
\date{\today}

\section{Introduction}

Recent measurements of the azimuthal momentum distributions of
particles produced in high-multiplicity $p+Pb$ collisions at the LHC
have revealed large asymmetries $v_n=\langle\cos
n\phi\rangle$~\cite{pPb_vn}. Remarkably, the octupole asymmetry is of
the same order of magnitude as the quadrupole. In this paper, we argue
that these asymmetries in the final state could reflect a snapshot of
the fluctuations in the Pb target taken in the instant of the
collision.  Rotational symmetry is spontaneously broken (locally) by a
fluctuation in the target of arbitrary shape. Such random fluctuations
generically lead to large local azimuthal anisotropies in
coordinate-space~\cite{Avsar:2010rf,Bzdak:2013rya}. 

In this paper we propose an alternative to (almost) dissipationless hydrodynamic expansion in pA
collisions~\cite{Bozek,Shuryak:2013ke} for the conversion of
coordinate-space fluctuations into asymmetries in momentum
space. Here, non-zero $v_n$'s emerge due to the orientation of a
projectile ``dipole'' relative to the global orientation of the event
determined by a fluctuation in the target. We show that parity even
moments $v_{2n}$ are generated by the real part of the dipole forward
scattering amplitude while parity odd moments $v_{2n+1}$ arise from
its imaginary part. While this argument for the generation of even and odd $v_n$'s is quite general, the actual calculation of these moments directly from QCD is quite challenging. In fact, calculations of $v_2$ in proton-nucleus collisions have been
carried out within the ``Color Glass Condensate'' approach, see~\cite{Venugopalan:2013cga} for a recent review, though at present it is not clear if a large $v_3$ emerges as well. 

In this aspect, the holographic correspondence \cite{Maldacena:1997re} may provide an interesting alternative where calculations can be done in a non-perturbative manner in strongly-coupled non-Abelian gauge theories using a (higher dimensional) effective theory in the semi-classical approximation which includes gravity (among other fields). In the holographic approach, the real and imaginary parts of the dipole forward scattering amplitude in a proton-nucleus collision can be extracted by studying a classical string described by the Nambu-Goto action (which corresponds to the probe dipole) coupled to the underlying nontrivial background fields (such as the metric, the dilaton, the Neveu-Schwarz 2-form and etc) that give support to a single holographic shockwave solution, used here to model the traveling nucleus. We outline such calculation below and show how parity even angular moments $v_{2n}$ are generated by the real part of the light-like Wilson loop due to the contribution from the background metric to the Nambu-Goto action while parity odd moments $v_{2n+1}$ arise from the imaginary part of
the light-like Wilson loop, which is naturally induced by a fluctuating Neveu-Schwarz 2-form.

Our treatment is distinct from that of Ref.~\cite{wSW-sSW} where a
proton-nucleus collision is modeled as an asymmetric collision of holographic
shockwaves \cite{Janik:2005zt}, or from ${\cal R}$-current deep
inelastic scattering~\cite{Hatta:2007he}. In our approach the gauge-gravity
correspondence is used only to evaluate the ``target field averages''
such as the light-like Wilson loop, analogous to the computation of
the jet quenching parameter $\hat{q}$ in ref.~\cite{Liu:2006ug}. We do
not use it to describe the actual collision; thus, no black hole is
formed and the projectile quark is not stopped (a non-stopping solution in a holographic
framework has indeed been found for thin enough projectile and target in \cite{Casalderrey-Solana:2013aba}). The form of the
scattering amplitude of a projectile quark with large light-cone
momentum is taken from QCD.

This paper is organized as follows. In the next section we briefly review how the quark-nucleus elastic scattering is described within the hybrid formalism and give the general argument which shows how nonzero $v_n$'s may be generated by a projectile dipole scattering off fluctuations in the target nucleus with a size on the order of the dipole. In Section \ref{holo} perform the holographic calculation of the $v_n$'s. Our conclusions and outlook can be found in Section \ref{conclusions}.

\section{Quark-Nucleus elastic scattering in the hybrid formalism}\label{hybrid}

In the so-called hybrid formalism the proton projectile is treated as
a beam of collinear partons with a large light-cone momentum $p^-$
which probe the field of the target. At large Feynman-$x_F$ the
contribution from quarks dominates while particles with $x_F\ll 1$ are
mainly gluons. We describe here mainly the case of quark scattering
but it is straightforward to obtain the contribution from gluons by
considering the scattering amplitude in the adjoint
representation. The main difference, however, is that the forward
scattering amplitude for an adjoint dipole is real and thus can only
generate parity even moments $v_{2n}$.

Our approach is based on the intuitive picture that a high-energy
projectile parton couples weakly to the target field. However, over
some intermediate range of semi-hard transverse momenta the target is
modeled as a holographic shockwave before it turns into a beam of
non-interacting asymptotically free partons at very high transverse
momentum $p_\perp$ (far exceeding its saturation scale $Q_s$ computed
below). Whether or not the proposed picture has a viable theoretical
justification remains to be seen; it is not clear if the limit of
eikonal, recoil less propagation of a high-energy projectile parton is
well defined in the present context~\cite{YK_eikonal}. We shall follow
\cite{Liu:2006ug} and assume the validity of this eikonal description
in this paper.

If we assume that to leading order in $p_\perp/p^-$ projectile partons
propagate on eikonal trajectories then the amplitude corresponding
to elastic scattering from momentum $p$ to $q$ is~\cite{BjKS}
\bea
\left< {\rm out}, q| {\rm in}, p\right> &\equiv& \bar{u}(q) \tau(q,p)
u(p) \\
\tau(q,p) &=& 2\pi \delta(p^--q^-) \; \gamma^- \int d^2\vec{x} \left[
  V(\vec{x})-1\right] e^{i (\vec{p}-\vec{q})\cdot \vec{x}}~.
\eea
Here,
\be
V(\vec{x}) = {\cal P} \exp \left[ ig \int dx^- A^+(x^-,\vec{x})\right]
\ee
is a Wilson line along the light cone. Upon squaring the
amplitude \cite{comment1} the scattering cross section can be
written as \cite{Dumitru:2002qt}
\be \label{eq:qA_el}
\frac{d\sigma_{qA}}{d^2\vec{b} \,d^2\vec{q}} = \frac{1}{(2\pi)^2}
\int d^2\vec{x}\, e^{-i\vec{q}\cdot \vec{x}}\,
\left(\left< \frac{1}{N_c} {\rm tr}\,\left(
W(\vec{x},\vec{b})
-V(\vec{b}-\vec{x}/2)-V^\dagger(\vec{b}+\vec{x}/2)
\right) \right>+1\right)~.
\ee
Here, $\vec{b}$ denotes the impact parameter of the collision and
$W(\vec{x},\vec{b})$ is a light-like Wilson loop of width given by
$|\vec{x}|$. In covariant gauge
$W(\vec{x},\vec{b})=V^\dagger(\vec{b}+\vec{x}/2) \, V(\vec{b}-\vec{x}/2)$
and this is commonly referred to as the dipole unintegrated gluon
distribution~\cite{MuellerDipole}. The size of the dipole corresponds
to the shift of the transverse coordinate of the eikonal quark line
from the amplitude to the complex conjugate amplitude, respectively.

Thus, the quark-nucleus cross section is written as a Fourier transform
of the dipole S-matrix,
\be
\frac{d\sigma_{qA}}{d^2\vec{b} \,d^2\vec{q}} = \frac{1}{(2\pi)^2}
\int d^2\vec{x}\, e^{-i\vec{q}\cdot \vec{x}}\, S(\vec{x}) =
\frac{1}{(2\pi)^2}
\int d^2\vec{x}\, e^{-i\vec{q}\cdot \vec{x}} \left(
D(\vec{x}) + iO(\vec{x})
\right)~.
\ee
Here, $D(\vec{x}) = {\rm Re}~S(\vec{x})$ and $O(\vec{x}) = {\rm
  Im}~S(\vec{x})$ denote the real and imaginary parts of the S-matrix,
respectively. Since the left hand side of this equation is manifestly
real, we must have that $D(\vec{x})=D(-\vec{x})$ is even under
exchange of the quark and anti-quark lines, while
$O(\vec{x})=-O(-\vec{x})$ is odd. It follows that $D(\vec{x})$ is
responsible for generating non-zero $v_{2n}=\langle\cos n\phi\rangle$
which are even under $\phi\to\phi+\pi$ (resp.\ $\vec{x}\to-\vec{x}$).
On the other hand, $v_{2n+1}$ is odd under $\phi\to\phi+\pi$ and hence
can only arise from $O(\vec{x})$.

Eq.\ (\ref{eq:qA_el}) can be turned into a physical $pA\to h+X$ single
inclusive cross section for production of a hadron of type $h$ via a
convolution with a proton-parton distribution and a corresponding
$q\to h$ fragmentation
function~\cite{Dumitru:2005gt,Altinoluk:2011qy,Chirilli:2011km}. Here
we will only need Eq.~(\ref{eq:qA_el}). Below, we employ the
holographic correspondence \cite{Maldacena:1997re} to compute the
light-like Wilson loop $W(\vec{x},\vec{b})$ in the field of a shockwave in strongly-coupled
$\mathcal{N}=4$ Supersymmetric Yang-Mills (SYM) theory with a large
number of colors, $N_c$. The essential point is to consider
scattering of a dipole whose angular orientation couples to
fluctuations in the target.

\section{Light-like Wilson loop in a holographic shockwave background}\label{holo}

The nucleus is traveling along the $x^+$ axis with a light-cone
momentum $p^+$. We are interested in holographic shockwave solutions of the form
\begin{equation}
\langle \hat{T}_{--}(x^-,\vec{x})\rangle=\frac{N_c^2}{2\pi^2}\,
p^+\,\delta(x^-)\,\mu^2\,f(\vec{x})
\label{shockwave}
\end{equation}
where $f(\vec{x})$ describes the energy density distribution of the
target in the transverse plane, and $\mu^2 \sim A^{1/3}$
\cite{Albacete:2008ze,Taliotis:2009ne} is the transverse density scale of the
shockwave. This non-uniform shockwave at the boundary can be obtained
from a source $J(\vec{x})\delta(z-1/\mu)$ in the bulk convoluted with
the Green's function found in \cite{Gubser:2008pc}, \be
f(\vec{x})=\int
d^2\vec{x}\,'\,\frac{J(\vec{x}\,')}{\left(1+\mu^2|\vec{x}-\vec{x}\,'|^2\right)^3
}\,.
\ee The metric is a solution of the 5 dimensional Einstein's equations
with a negative cosmological constant $\sim 1/L^2$ with a source
\cite{Gubser:2008pc,Kovchegov:2009du}. It has the form
\begin{eqnarray}
ds^2 &=&
\frac{L^2}{z^2}\left(p^+\delta(x^-)\mu^2\mathcal{F}(z,\vec{x})
z^4\,dx^{-2} - 2 dx^+dx^-   + d\vec{x}^2+dz^2\right)~,
\label{geometry}
\end{eqnarray}
where $\lim_{z\to 0}\mathcal{F}(z,\vec{x})=f(\vec{x})$. We note that
ref.~\cite{Beuf:2009mk} found a family of regular shockwave solutions
which, in the homogeneous limit, have vanishing $\langle
\hat{T}_{--}(x^-,\vec{x})\rangle$. In fact, those bulk solutions do
not approach $\sim z^4$ near the boundary and are, thus, qualitatively
distinct from the type of solutions found in
refs.~\cite{Janik:2005zt,Gubser:2008pc}. In this paper, we consider
that the nucleus is, on average, essentially uniform in the transverse
plane over scales probed by the dipole and, thus, we shall restrict to
solutions of the form~(\ref{geometry}).

One can now compute the light-like Wilson loop in this background
\cite{commentloop}. The rectangular loop $\mathcal{C}$ is defined on
the $x^-$ axis with transverse size $\vec{d}$ and it is the boundary
for a minimal surface in the bulk
\cite{Maldacena:1998im,comment0}. When the radius of AdS$_5$ is much
larger than the string length $\ell_s$, i.e., $L^2/\alpha' \gg 1$,
where $\alpha' = \ell_s^2$, this is obtained by minimizing the
Nambu-Goto action
\begin{equation}
S_{NG} = \frac{1}{2\pi \alpha'} \int_{\Omega} d^2\sigma \,\sqrt{-{\rm
    det}\,h_{ab}}\, 
\label{nambugoto}
\end{equation}
where $\Omega$ denotes the worldsheet, $h_{ab} = g_{MN}\partial_a X^M
\partial_b X^N$ is the worldsheet metric, and $X^M(\sigma)$ is the
embedding function that describes the string worldsheet in the
bulk. In principle, the effective action for the string should also
include the coupling to other background fields such as the dilaton
and the Neveu-Schwarz (NS) 2-form \cite{polchinski} from the NS-NS
sector but these are taken to be either vanishing or pure gauge (more
on that below).

Our calculation for the light-like Wilson loop closely follows
the one performed in \cite{Liu:2006ug} to obtain the jet quenching
parameter $\hat{q}$ coefficient in the strongly coupled
$\mathcal{N}=4$ SYM plasma. However, for our shockwave the gauge
theory is not at finite temperature and, thus, the background geometry
does not have a horizon.  The light-like Wilson loop is given in terms
of the on-shell action as $\langle {\rm tr}\,
W(\mathcal{C})\rangle/N_c= e^{iS_{on-shell\,NG}}$.

The string worldsheet coordinates are $\tau=x^-$ and $\sigma$ and the
worldsheet embedding function is $X^M =
(x^-,\vec{b}+\sigma\vec{d},0,d\,u(\sigma))$ with $\sigma \in
(-1/2,1/2)$. The endpoints of the string are located at the boundary
at $\vec{b}-\vec{d}/2$ and $\vec{b}+\vec{d}/2$. With the metric
\eqref{geometry} the Nambu-Goto action becomes
\begin{eqnarray}
iS_{NG} &=& -\frac{\sqrt{\lambda}}{2\pi} \,A^{1/6} \mu
\,d\,\int_{-1/2}^{1/2}d\sigma\,\sqrt{\mathcal{F}\left(d\,u(\sigma),\vec{b}+\sigma
  \vec{d}\,\right)}\sqrt{1+u'(\sigma)^2}\,.
\label{lightlikeaction}
\end{eqnarray}
The explicit factor of $A^{1/6}$ arises from the integration over the
longitudinal coordinate $x^-$~\cite{Albacete:2008ze,Taliotis:2009ne,delta_x-}.

Due to the properties of the function $\mathcal{F}$, the integrand of
the action is finite at the boundary, $u\to 0$, as opposed to the case
of a time-like rectangular Wilson loop in vacuum
\cite{Maldacena:1998im} where it diverges as $\sim 1/u^2$. Therefore,
any configuration where $u'(\sigma)\neq 0$ will necessarily increase
the worldsheet area. Hence, the minimal surface must be the one in
which the string remains at the boundary for all $\sigma$, i.e., the
string does not fall into the bulk. In fact, $u(\sigma)=0$ is clearly
a solution of the equations of motion that satisfies the boundary
conditions, which is consistent with the fact that a light-like string
configuration costs zero energy in AdS$_5$ \cite{Ito:2007zy}.

Therefore, the on-shell action is obtained by setting $u'(\sigma)=0$
and $u(\sigma)=0$ which leads to
\begin{eqnarray}
iS_{on-shell\,NG} &=& -\frac{\sqrt{\lambda}}{2\pi} \,A^{1/6} \mu\, d\,
\int_{0}^{1/2}d\sigma\,\left(\sqrt{f\left(\vec{b}+\sigma
  \vec{d}\,\right)}+\sqrt{f\left(\vec{b}-\sigma
  \vec{d}\,\right)}\right)\,.
\label{lightlikeactiononshell}
\end{eqnarray}
Note the $\vec{d}\to-\vec{d}$ symmetry and the fact that $iS_{NG}$ is
real. In what follows we assume that the nucleus is much larger than
the dipole and that its density over large distance scales is
homogeneous. Thus, we can set $\vec{b}=0$. Furthermore, in the absence
of fluctuations we have $f=1$ and so the forward dipole scattering
amplitude becomes
\be
W(d) = e^{-\frac{\sqrt{\lambda}}{2\pi} A^{1/6}\mu\,d}~.
\ee
The ``saturation scale'' where $W\sim e^{-1}$ therefore is
\be
Q_s = \frac{\sqrt{\lambda}}{2\pi} A^{1/6} \mu~.
\ee
$Q_s$ obtained by averaging the Wilson loop in a shockwave background
increases very rapidly with the thickness of the nucleus, $Q_s\sim
A^{1/3}$~\cite{Albacete:2008ze}.

The transverse momentum distribution of scattered quarks is
\be \label{eq:pt-distribution}
\frac{d\sigma_{qA}}{d^2\vec{b} \,d^2\vec{q}_\perp} =
\frac{Q_s}{(Q_s^2+p_\perp^2)^{3/2}}~.
\ee
We should stress that this result is not supposed to apply at very
large $p_\perp$ where from perturbative QCD $d\sigma_{qA}/d^2p_\perp
\sim \alpha_s^2/p_\perp^4$~\cite{ptUV}. However, in the non-linear regime at
intermediate $p_T$ one does indeed expect a ``flatter'' transverse
momentum distribution similar to~(\ref{eq:pt-distribution}).

\subsection{Fluctuations in the holographic shockwave and even moments
  $v_{2n}$}

We introduce fluctuations of the density of the shockwave in terms of
their Fourier spectrum,
\bea
f\left(\vec{b}+\sigma \vec{d}\right) &=& 1 + \delta
f(\vec{b}+\sigma \vec{d}) \\
\delta f(\vec{x}) &=& \int \frac{d^2k}{(2\pi)^2} \, \delta f(\vec{k})
\, e^{i \vec{k} \cdot \vec{x}}~.  \label{eq:Fourier_df}
\eea
$\delta f$ describes ``classical'' fluctuations in the target which
contribute $\sim N_c^2$ to the energy-momentum tensor,
c.f.\ eq.~(\ref{shockwave}). For simplicity here we assume that 
\be \delta f(\vec{k}) =
\frac{1}{2}(2\pi)^2 \, \mathcal{A} (1/|\vec{k}_0|) \left[\delta
  (\vec{k} -\vec{k}_0) + \delta (\vec{k} +\vec{k}_0)+i(\delta (\vec{k}
  -\vec{k}_0) - \delta (\vec{k} +\vec{k}_0)) \right]\,, 
\ee 
i.e., that the fluctuation is dominated by a single wave number and
direction though one could also average over some suitable
distribution. $\mathcal{A} (1/|\vec{k}_0|)$ is the amplitude of the
fluctuation at the scale $k_0$; we shall denote the typical length
scale $1/|\vec{k}_0|$ of fluctuations as $\ell$, and the azimuthal
orientation of the dipole as $\phi$ so that $\vec{d}\cdot\vec{k}_0 =
d/\ell \cos\phi$. Note that in order to obtain $v_n$ one only averages
over this relative angle $\phi$ while the global orientation is fixed;
alternatively, the moments could be defined from two-particle
cumulants~\cite{vn_cumulant}
\be
v_n^2 = \left< e^{in(\phi_1-\phi_2)} \right>~,
\ee
where $\phi_1$ and $\phi_2$ are the azimuthal angles of any two
particles from the same event.

Expanding the square root in (\ref{lightlikeactiononshell}) for small
amplitude fluctuations we find
\begin{eqnarray}
iS_{on-shell} &=& -Q_s \,d\,\left[1+\mathcal{A}(\ell)
\frac{\sin\left(\frac{d}{2\ell}\,
\cos\,\phi\right)}{\frac{d}{\ell}\cos\,\phi}\right]\,.
\label{lightlikeactiononshelleven}
\end{eqnarray}

The fluctuations generate asymmetries for the
multipole moments of the $p_\perp$ distribution,
\be
\frac{d\sigma_{qA}}{d^2\vec{b} \; p_\perp dp_\perp d\phi_p} =
\frac{1}{(2\pi)^2}
\int x_\perp dx_\perp d\phi\, e^{-ip_\perp x_\perp\cos(\phi-\phi_p)}\,
e^{iS_{on-shell\,NG}}~\,.
\label{distribution}
\ee
where we denoted the transverse size of the dipole as
$\vec{x}_\perp$. Parametrically, nonzero moments of this distribution
will be of order $\sim Q_s \,\ell\, \mathcal{A}(\ell)$. However, the
action~(\ref{lightlikeactiononshelleven}) is even under
$\phi\to\phi+\pi$ and, thus, it can only generate even moments of the
angular distribution.

\subsection{Fluctuations of the NS 2-form and odd moments $v_{2n+1}$'s}

Odd moments of the angular distribution can be generated in this
approach from the imaginary part of the dipole-nucleus $S$-matrix
$\langle V^\dagger(\vec{x}) V(\vec{y})\rangle$ which includes $C$-even
``pomeron'' and $C$-odd ``odderon'' exchanges. Projecting on odd-even
in-out states, the latter corresponds to the imaginary part ${\Im}m\,
S(\vec{x},\vec{y}) = \langle O(\vec{x},\vec{y})\rangle$ where the
odderon operator is given by~\cite{odderon}
\be
O(\vec{x},\vec{y}) = \frac{1}{2i\, N_c} \; {\rm tr}\,\left(
V^\dagger(\vec{x})V(\vec{y}) -
V^\dagger(\vec{y})V(\vec{x}) \right)~.
\ee
The odderon has been identified with the fluctuations of the
anti-symmetric Neveu-Schwarz-Neveu-Schwarz (NS-NS) 2-form $B_{MN}$ in
the bulk \cite{Brower:2008cy,Avsar:2009hc}. In the dual holographic
description used here, the contribution to the effective action of the string that is odd
under the $\vec{d} \to -\vec{d}$ operation should arise from the coupling
of the NS 2-form field to the string in the Nambu-Goto
action
\be
S_{NS-NS} = \frac{1}{4\pi \alpha'}\int_{\Omega}
d^2\sigma\,B_{MN}\,\epsilon_{ab}\,\partial_a X^M \partial_b X^N 
\label{KRaction}
\ee
where $\epsilon_{ab}$ is the Levi-Civita symbol on the worldsheet
\cite{commentKR}. For a single shockwave we consider a pure gauge
NS-field $B_{MN}$ which does not alter the equations of motion of
supergravity \cite{polchinski} so that the solutions for the background metric and for
the string remain valid. [After a collision of two shockwaves this is no
  longer the case, just as the metric is no longer of the
  form~(\ref{geometry}).]

Contributions such as~(\ref{KRaction}) should indeed lead to parity
odd moments of the angular distribution (\ref{distribution}).
Choosing a gauge where $B_{-M}=0$, this term in the action
becomes (using the worldsheet embedding defined earlier)
\be
S_{NS-NS} = \frac{1}{2\pi \alpha'}\int_{-\infty}^{\infty}dx^-\,
\int_0^{1/2}d\sigma\, \,\vec{d}\cdot \left[ \vec{B}_{-\vec{x}_\perp}
  (x^-,\vec{b} + \vec{d}\sigma,0)+ \vec{B}_{-\vec{x}_\perp}
  (x^-,\vec{b}-\vec{d}\sigma,0)\right]\,.
\ee
This action is purely real. Thus, it contributes a phase to the total amplitude
$\exp\{iS_{NG}+iS_{NS-NS}\}$ which is odd under $\vec{d}\to
-\vec{d}$. Therefore, in this more general scenario, odd moments for
the angular distribution such as $v_3$ should be nonzero and, again,
of order $\ell\, Q_s$. Specifically, terms such as
\be \label{eq:d*nabla_f}
i \int \limits_{-1/2}^{1/2}
d\sigma\,  \vec{d}\cdot\vec\nabla\, f(\sigma\vec{d}) = -2
i{\cal A}(\ell)\;\sin\left(\frac{d}{2\ell}\cos\phi\right)~,
\ee
can arise. Indeed, this is parity ($\phi\to\phi+\pi$) odd and
generates odd $v_n$'s up to $n\sim d/\ell$.

\section{Conclusions}\label{conclusions}

In summary, we have argued that azimuthal asymmetries $v_n$ in p+A
collisions may arise from scattering of a dipole on random
fluctuations in the target; the fluctuations are assumed to be
``classical'' so that $\delta T_{--}\sim N_c^2$.

The real (imaginary) part of the dipole-nucleus S-matrix is even (odd)
under exchange of the quark and anti-quark lines corresponding to
charge conjugation of the Wilson loop, and gives rise to parity even
(odd) angular moments $v_n$. This is a simple and quite general
mechanism that allows for the generation of nonzero Fourier moments of
hadron yields in proton-nucleus collisions. Whether or not this is
indeed the main effect behind the nonzero $v_n$'s (in particular, of $v_3$)
in these collisions still remains to be verified.

We have used the holographic correspondence to determine the
properties of light-like Wilson loops in a shockwave background in
strongly-coupled $\mathcal{N}=4$ SYM. We use this as a toy model for
the actual calculation of $v_n$'s in QCD. In the holographic
description the contribution from the metric to the Nambu-Goto action
produces parity even distributions while the coupling of a fluctuating
NS-NS 2-form field with the classical string can generate odd
moments. More detailed numerical calculations of $v_n(p_\perp)$ could
potentially provide information on the spectrum of fluctuations, such
as if there is a dominant length scale and amplitude (as assumed here,
for simplicity).

It would be interesting to generalize the calculation performed here
to take into account other effects such as the presence of a confining
scale in shockwave solutions \cite{Kiritsis:2011yn}. This requires
different shockwave solutions such that the string connecting the
sources does sag into the bulk. We leave this to a future study.

\section*{Acknowledgements}
J.~N.\ thanks Conselho Nacional de Desenvolvimento Cient\'ifico e
Tecnol\'ogico (CNPq) and Funda\c c\~ao de Amparo \`a Pesquisa do
Estado de S\~ao Paulo (FAPESP) for support. A.~D. thanks the
University of S\~ao Paulo for their hospitality during a visit when
this work was initiated, and gratefully acknowledges support by the
DOE Office of Nuclear Physics through Grant No.\ DE-FG02-09ER41620 and
from The City University of New York through the PSC-CUNY Research
Award Program, grant 66514-0044. The authors thank Y.~Hatta for
discussions at the Yukawa Institute for Theoretical Physics, Kyoto
University, where this work was completed during the YITP-T-13-05
workshop on ``New Frontiers in QCD''. We thank W.~van der Schee,
H.~Nastase, and A.~Taliotis for comments and Y.~Kovchegov for a
critical reading of the manuscript before publication and for helpful
comments about the odderon and the eikonal limit in QCD.




\begin{thebibliography}{99}

\bibitem{pPb_vn}
S.~Chatrchyan {\it et al.}  [CMS Collaboration],
Phys.\ Lett.\ B {\bf 718}, 795 (2013);
B.~Abelev {\it et al.}  [ALICE Collaboration],
Phys.\ Lett.\ B {\bf 719}, 29 (2013);
G.~Aad {\it et al.}  [ATLAS Collaboration],
Phys.\ Rev.\ Lett.\  {\bf 110}, 182302 (2013);
G.~Aad {\it et al.}  [ATLAS Collaboration],
Phys.\ Lett.\ B {\bf 725}, 60 (2013);
S.~Chatrchyan {\it et al.}  [CMS Collaboration],
Phys.\ Lett.\ B {\bf 724}, 213 (2013).

\bibitem{Avsar:2010rf} 
E.~Avsar, C.~Flensburg, Y.~Hatta, J.~-Y.~Ollitrault and T.~Ueda,
Phys.\ Lett.\ B {\bf 702}, 394 (2011).

\bibitem{Bzdak:2013rya} 
A.~Bzdak, P.~Bozek and L.~McLerran,
arXiv:1311.7325 [hep-ph].

\bibitem{Bozek}
P.~Bozek and W.~Broniowski,
Phys.\ Lett.\ B {\bf 718}, 1557 (2013);
P.~Bozek and W.~Broniowski,
Phys.\ Rev.\ C {\bf 88}, 014903 (2013).

\bibitem{Shuryak:2013ke} 
E.~Shuryak and I.~Zahed,
Phys.\ Rev.\ C {\bf 88}, 044915 (2013).

\bibitem{Venugopalan:2013cga} 
R.~Venugopalan,
arXiv:1312.0113 [hep-ph].

\bibitem{Maldacena:1997re} 
  J.~M.~Maldacena,
  Adv.\ Theor.\ Math.\ Phys.\  {\bf 2}, 231 (1998)
  [hep-th/9711200].





\bibitem{wSW-sSW}
J.~L.~Albacete, Y.~V.~Kovchegov and A.~Taliotis,
JHEP {\bf 0905}, 060 (2009);
J.~Casalderrey-Solana, M.~P.~Heller, D.~Mateos, W.~van~der~Schee,
arXiv:1312.2956.

\bibitem{Janik:2005zt} 
  R.~A.~Janik and R.~B.~Peschanski,
  Phys.\ Rev.\ D {\bf 73}, 045013 (2006)
  [hep-th/0512162].

\bibitem{Hatta:2007he} 
Y.~Hatta, E.~Iancu and A.~H.~Mueller,
JHEP {\bf 0801}, 026 (2008);
JHEP {\bf 0801}, 063 (2008).

\bibitem{Liu:2006ug} 
  H.~Liu, K.~Rajagopal and U.~A.~Wiedemann,
  Phys.\ Rev.\ Lett.\  {\bf 97}, 182301 (2006)
  [hep-ph/0605178]; 
  JHEP {\bf 0703}, 066 (2007)
  [hep-ph/0612168].
  
\bibitem{Casalderrey-Solana:2013aba} 
  J.~Casalderrey-Solana, M.~P.~Heller, D.~Mateos and W.~van der Schee,
  Phys.\  Rev.\  Lett.\  111, {\bf 181601} (2013)
  [Phys.\ Rev.\ Lett.\  {\bf 111}, 181601 (2013)]
  [arXiv:1305.4919 [hep-th]].
  
\bibitem{YK_eikonal}
Y.~Kovchegov, private communication.

\bibitem{BjKS}
J.~D.~Bjorken, J.~B.~Kogut and D.~E.~Soper,
Phys.\ Rev.\ D {\bf 3}, 1382 (1971).

\bibitem{comment1} The amplitude and its complex conjugate must be
  parallel transported to $\vec{x}=0$ unless $A^+$ is obtained in
  covariant gauge.

\bibitem{Dumitru:2002qt} 
A.~Dumitru and J.~Jalilian-Marian,
Phys.\ Rev.\ Lett.\  {\bf 89}, 022301 (2002).

\bibitem{MuellerDipole}
A.~H.~Mueller,
Nucl.\ Phys.\ B {\bf 415}, 373 (1994);
Nucl.\ Phys.\ B {\bf 437}, 107 (1995);
A.~H.~Mueller and B.~Patel,
Nucl.\ Phys.\ B {\bf 425}, 471 (1994).

\bibitem{Dumitru:2005gt} 
A.~Dumitru, A.~Hayashigaki and J.~Jalilian-Marian,
Nucl.\ Phys.\ A {\bf 765}, 464 (2006).

\bibitem{Altinoluk:2011qy} 
T.~Altinoluk and A.~Kovner,
Phys.\ Rev.\ D {\bf 83}, 105004 (2011).

\bibitem{Chirilli:2011km} 
G.~A.~Chirilli, B.-W.~Xiao and F.~Yuan,
Phys.\ Rev.\ Lett.\  {\bf 108}, 122301 (2012);
Phys.\ Rev.\ D {\bf 86}, 054005 (2012).


  
\bibitem{odderon}
Y.~V.~Kovchegov, L.~Szymanowski and S.~Wallon,
Phys.\ Lett.\ B {\bf 586}, 267 (2004);
Y.~Hatta, E.~Iancu, K.~Itakura and L.~McLerran,
Nucl.\ Phys.\ A {\bf 760}, 172 (2005);
S.~Jeon and R.~Venugopalan,
Phys.\ Rev.\ D {\bf 71}, 125003 (2005).



\bibitem{Albacete:2008ze} 
  J.~L.~Albacete, Y.~V.~Kovchegov and A.~Taliotis,
  JHEP {\bf 0807}, 074 (2008)
  [arXiv:0806.1484 [hep-th]].
  
\bibitem{Taliotis:2009ne} 
  A.~Taliotis,
  Nucl.\ Phys.\ A {\bf 830}, 299C (2009)
  [arXiv:0907.4204 [hep-th]].
  
\bibitem{Gubser:2008pc} 
  S.~S.~Gubser, S.~S.~Pufu and A.~Yarom,
  Phys.\ Rev.\ D {\bf 78}, 066014 (2008)
  [arXiv:0805.1551 [hep-th]].
  
\bibitem{Kovchegov:2009du} 
  Y.~V.~Kovchegov and S.~Lin,
  JHEP {\bf 1003}, 057 (2010)
  [arXiv:0911.4707 [hep-th]].
  
\bibitem{Beuf:2009mk} 
  G.~Beuf,
  Phys.\ Lett.\ B {\bf 686}, 55 (2010)
  [arXiv:0903.1047 [hep-th]].
  
\bibitem{Grigoryan:2011cn} 
  H.~R.~Grigoryan and Y.~V.~Kovchegov,
  Nucl.\ Phys.\ B {\bf 852}, 1 (2011)
  [arXiv:1105.2300 [hep-th]].


\bibitem{commentloop} Given the matter content of $\mathcal{N}=4$ SYM theory, the Wilson loop also contains the coupling to the six $SU(N_c)$ adjoint scalars $X^I$. We follow \cite{Grigoryan:2011cn} and consider an average over all the angles $\theta^I$ on $S^5$ to eliminate the dependence of the string worldsheet on modes that carry nonzero Kaluza-Klein charge. 

\bibitem{Maldacena:1998im} 
  J.~M.~Maldacena,
  Phys.\ Rev.\ Lett.\  {\bf 80}, 4859 (1998)
  [hep-th/9803002].



\bibitem{comment0} Wilson loops with light-like segments have been extensively studied within the gauge/gravity correspondence, see, e.g., L.~F.~Alday and J.~M.~Maldacena,
  JHEP {\bf 0706}, 064 (2007).
  
\bibitem{polchinski} J.~Polchinski, {\it String Theory}, Vol.\ 2,
  Cambridge University Press, 1998.


\bibitem{delta_x-}
To see this one writes $\delta(x^-)
  \to \frac{1}{a}\, \Theta(x^-)\;\Theta(a-x^-)$ with $a\sim A^{1/3}$.

\bibitem{Ito:2007zy} 
  K.~Ito, H.~Nastase and K.~Iwasaki,
  Prog.\ Theor.\ Phys.\  {\bf 120}, 99 (2008)
  [arXiv:0711.3532 [hep-th]].
  
\bibitem{ptUV}
A $\sim1/p_\perp^4$ behavior at high $p_\perp$ is required also in
order for $\langle p_\perp\rangle$ to be finite and independent of any
UV cutoffs. This is a necessary condition, even at the level of purely
elastic scattering, for the validity of eikonal projectile
trajectories in the limit $\langle p_\perp\rangle/p^- \to 0$.

\bibitem{vn_cumulant}
N.~Borghini, P.~M.~Dinh and J.~-Y.~Ollitrault,
Phys.\ Rev.\ C {\bf 63}, 054906 (2001).

\bibitem{Brower:2008cy} 
  R.~C.~Brower, M.~Djuric and C.~-ITan,
  JHEP {\bf 0907}, 063 (2009)
  [arXiv:0812.0354 [hep-th]].
  
\bibitem{Avsar:2009hc} 
  E.~Avsar, Y.~Hatta and T.~Matsuo,
  JHEP {\bf 1003}, 037 (2010)
  [arXiv:0912.3806 [hep-th]].

\bibitem{commentKR} The action in Eq.\ (\ref{KRaction}) has to be
  supplemented by the boundary term $\int_{\partial \Omega}A_1$ (where
  $A_M$ is a 1-form) to preserve the gauge invariance associated with
  $B_{MN}$. This term is subleading in $\alpha'$ and it will be
  omitted in this paper.



\bibitem{Kiritsis:2011yn} 
  E.~Kiritsis and A.~Taliotis,
  JHEP {\bf 1204}, 065 (2012)
  [arXiv:1111.1931 [hep-ph]].
  

  


\end{thebibliography}
\end{document}